\documentclass[10pt,aps,pra,twocolumn,showpacs,superscriptaddress,nofootinbib]{revtex4-2}
\usepackage[english]{babel}
\usepackage[usenames, dvipsnames]{color} 
\usepackage{graphicx} 
\usepackage{bm} 
\usepackage{amsmath} 
\usepackage{amsthm} 
\usepackage{amssymb} 
\usepackage{times} 
\usepackage[pdftex,colorlinks=true,urlcolor= blue,linkcolor=Blue,citecolor=RedViolet]{hyperref}
\usepackage{multirow}
\usepackage{graphicx}
\usepackage{bm}
\usepackage{physics}
\usepackage{amssymb,amsmath}
\usepackage[T1]{fontenc}
\usepackage[utf8]{inputenc}
\usepackage{lmodern}
\usepackage{stackrel}
\usepackage{bbold}

\usepackage{ulem}

\newcommand{\aspasc}{\textquotedblleft}
\newcommand{\aspasd}{\textquotedblright\ }

\usepackage[small,bf,justification=Justified]{caption}
\newcommand{\com}[1]{} 

\usepackage{soul}

\begin{document}

	\title{Benchmarking Variational Quantum Eigensolvers for Entanglement Detection in Many-Body Hamiltonian Ground States}

	\author{Alexandre Drinko}
	\email{adrinko@usp.br}
	\author{Guilherme I. Correr}
	\affiliation{Instituto de Física de São Carlos, Universidade de São Paulo, CP 369, 13560-970 São Carlos, Brazil}
	\author{Ivan Medina}
	\affiliation{Instituto de Física de São Carlos, Universidade de São Paulo, CP 369, 13560-970 São Carlos, Brazil}
	\affiliation{School of Physics, Trinity College Dublin, Dublin 2, Ireland}
	\author{Pedro C. Azado}
	\affiliation{Instituto de Física de São Carlos, Universidade de São Paulo, CP 369, 13560-970 São Carlos, Brazil}
	\author{Askery Canabarro}
	\affiliation{Instituto de Física de São Carlos, Universidade de São Paulo, CP 369, 13560-970 São Carlos, Brazil}
	\affiliation{Grupo de Física da Matéria Condensada, Núcleo de Ciências Exatas - NCEx, Campus Arapiraca, Universidade Federal de Alagoas, 57309-005 Arapiraca, Alagoas, Brazil}
	\affiliation{Department of Physics, Harvard University, Harvard University, Cambridge, Massachusetts 02138, USA}
	\affiliation{Quantum Research Center, Technology Innovation Institute, Abu Dhabi, UAE}
	
	\author{Diogo O. Soares-Pinto}
	\email{dosp@ifsc.usp.br}
	\affiliation{Instituto de Física de São Carlos, Universidade de São Paulo, CP 369, 13560-970 São Carlos, Brazil}
	
	\begin{abstract}
		Variational quantum algorithms (VQAs) have emerged in recent years as a promise to
		obtain quantum advantage. These task-oriented algorithms work in a hybrid loop combining a quantum processor and classical optimization.
		Using a specific class of VQA named variational quantum eigensolvers (VQEs), we choose some parameterized quantum circuits to benchmark them at entanglement witnessing and entangled ground state detection for many-body systems described by Heisenberg Hamiltonian, varying the number of qubits and shots. 
		Quantum circuits whose structure is inspired by the Hamiltonian interactions presented better results on cost function estimation than problem-agnostic circuits.
	\end{abstract}
	
	\maketitle

	\section{Introduction}
	Variational quantum algorithms (VQAs) have been noticed as an interesting promise to achieve quantum advantage in
	the Noisy Intermediate Scale Quantum (NISQ) era of
	quantum computation \cite{preskill2018quantum, leymann2020bitter,
		RevModPhys.94.015004,bharti2022noisy}.
	Under the restrictions on
	the number of qubits and amount of noise in the contemporary quantum computers,
	quantum computing tasks that demand less qubits and low circuit depth are
	interesting to explore the limits of the available hardware \cite{callison2022hybridalgorithms}.
	
	The VQAs are hybrid task-oriented algorithms that combine quantum and classical computation to estimate a cost (or objective) function encoding the solution of a specific problem. While the quantum computer deals with the system state preparation, transformation, and measurements, the classical computer works with the outputs,
	post-processing a adjustable set of parameters imprinted in the system state by a parameterized
	quantum circuit \cite{cerezo2021variational,cerezo2021cost}.  The chosen architecture of the circuit is usually termed \textit{Ansatz} in this context and specifies the sequence of gates applied on the qubits. This hybrid loop is then repeated to optimize the chosen cost function \cite{leymann2020bitter,cerezo2021variational,	callison2022hybridalgorithms,HEA_proposal,HVA_refprincipal,McClean_Ucoupledcluster}. 
	
	VQAs have been applied to solve a plethora of different tasks. In materials science, biology, and chemistry areas \cite{bauer2020quantum},  they have been used to optimize molecular geometry \cite{delgado2021variational} or find the ground state of molecular Hamiltonians \cite{HEA_proposal}, for example. VQAs also turned out to be an interesting tool for thermodynamics protocols, where we can mention the optimization of work extraction from quantum batteries \cite{medina2023vqeinspired,VQErg}, preparation of thermal states \cite{verdon2019quantum, consiglio2023variational,warren2022adaptive}, simulation of thermodynamic properties in metals \cite{silva2024simulating} and study of many body features such as phase transitions \cite{uvarov2020machine}. Interestingly enough, VQAs also find applications outside of the natural sciences ground. In the realm of the financial market, for instance, VQAs have been used to solve problems involving portfolio optimization \cite{canabarro2022quantum,wang2024variational,brandhofer2022qaoaportfolio}. Other applications where VQAs are used alongside other classical and quantum machine learning methods can be found in \cite{liu2023practical,havlivcek2019supervised,biamonte2017quantum,farhi2014qaoa,larocca2022groupinvqml,lins24}.  
	
	Many of the first proposals for VQAs were built from inspiration in the
	problems of searching for ground states of molecules \cite{McClean_Ucoupledcluster, Mcclean_nature_VQE_photonic}. This is a specific class of VQA, where the cost function
	takes the expectation value of the system's  Hamiltonian, named as Variational Quantum Eigensolver (VQE) \cite{tilly2022variational,cerezo2022stateeigensolver}.
	The VQE constitutes a powerful tool for studying the spectrum of interacting Hamiltonians \cite{higgott2019variational,jones2019variational,liu2019vqe}, and maybe its most prominent application is the estimation of Hamiltonians ground states  \cite{tilly2022variational,cerezo2022stateeigensolver},  state preparations \cite{kardashin2020certified} and also entanglement detection in many body systems \cite{consiglio2022variational}. It is well known that entanglement constitutes an important resource for quantum computation \cite{plenio2005introduction,horodecki2009quantum}, quantum
	metrology \cite{kok2004quantum} and quantum cryptography \cite{bruss2002optimal}. For bipartite quantum systems, there is a well-established approach to entanglement detection and quantification \cite{terhal2000bell,horodecki1996teleportation}.
	However, for multipartite systems, methods for entanglement verification becomes a challenging problem where no unique manner to classify and quantify it is currently available \cite{friis2019entanglement}. For this reason, 
	methods for entanglement detection and quantification are of utmost importance \cite{shahandeh2014structural,weilenmann2020entanglement}. Given this challenge, defining entanglement witnesses provides a practical method for detecting entanglement in experimental realizations \cite{amaro2020design,eisert2007quantitative,wang2024probing}.
	
	In particular, the detection of entangled ground states in many body systems is an active topic of research that finds applications in a multitude of protocols \cite{dowling2004energy,guhne2009entanglement,amaro2020design}. In this work, we use the VQE to optimize an entanglement witness and detect entanglement in the ground state of the Heisenberg Hamiltonian, which is an important Hamiltonian model for many body systems \cite{yamamoto1998elementary,vznidarivc2008many}. Due to the availability of different quantum hardware where this task could be performed, each one having a different set of gates available for building the Ansatz, the main goal of our work is to benchmark some parameterized quantum circuits. To attain this task, we carefully analyze the performance of entanglement detection for each Ansatz by varying the number of qubits in the Heisenberg Hamiltonian and also the number of shots, which represents the number of measurement outcomes available to stochastically construct the cost function.
	
	This work is organized as follows:
	In Sec. \ref{Sec.EW} we show the entanglement witness and the Hamiltonian
	used in all the simulations.
	In Sec. \ref{Sec.VQE} a brief review of variational quantum
	eigensolvers is presented, followed by the Sec. \ref{Sec.Ansatze} where the parameterized quantum
	circuits are chosen for the benchmark.
	The results are presented in Sec. \ref{Sec.Results} where we compare
	the optimization process for different numbers of shots, the analysis of the entanglement
	detection and then the ground state convergence, highlighting the differences 
	among the Ansätze and number of qubits.
	Finally in Sec. \ref{Sec.Discuss} the discussion of some important aspects
	about the benchmarking and conclusions of the work.
	
	\section{Hamiltonian as Entanglement Witness}\label{Sec.EW}
	To detect the existence of entanglement we can use an
	entanglement witness, $Z_{EW}$, that represents
	a sufficient, but not necessary, criterion to detect 
	entanglement \cite{dowling2004energy,horodecki2009quantum,guhne2009entanglement,friis2019entanglement}.
	For each entangled state, exists at least one entanglement witness capable of detecting it \cite{amico2008entanglement}.
	
	An entanglement witness $Z_{EW}$ is a Hermitian operator 
	that gives a positive expectation value for all separable states
	and a negative for at least one entangled state:
	\begin{align}
		\Tr\{Z_{EW} \rho\}< 0,& \text{\phantom{12} for at least one entangled $\rho$,} \label{eq:ew cond.1}\\
		\Tr\{Z_{EW} \rho\}\geq 0,& \text{\phantom{12} for all separable $\rho$.} \label{eq:ew cond.2}
	\end{align}
	
	There is a plethora of methods to implement an entanglement witness
	by considering different criteria \cite{guhne2009entanglement,friis2019entanglement}.
	We focus in Ref. \cite{dowling2004energy}, where it is presented how to
	witness entanglement in many body systems using the Hamiltonian as the 
	entanglement witness.
	The separable states are in a convex set $\mathcal{S}$, such that there
	exists a minimum separable energy $E_{sep}$ that corresponds to the
	lowest possible energy that a separable state can achieve,
	\begin{align}
		E_{sep}=\min_{\rho_{sep}\in\mathcal{S}}\Tr{\rho_{sep}H}.
	\end{align}
	If the ground state energy $E_0$ has a lower value than $E_{sep}$, we can define an entanglement gap $G=E_{sep}-E_0$, where all states for which the mean energy is inside the gap G are entangled.
	Then a Hamiltonian $H$ has a
	non-zero entanglement gap if and only if no ground state
	of $H$ is separable \cite{dowling2004energy}.
	
	In this sense, every Hamiltonian $H$ with a positive entanglement gap defines
	an energy-based entanglement witness
	\begin{align}
		Z_{EW}\equiv H-E_{sep}\mathbb{1}, \label{eq:ent.witness}
	\end{align}
	that satisfies the Eqs. (\ref{eq:ew cond.1}) and (\ref{eq:ew cond.2})
	for a set of entangled states bounded by the entanglement gap.
	
	It is important to remark that the entanglement witness represents a
	sufficient but not necessary condition to detect entanglement. Then, if the 
	witness application in a state results in a positive result, 
	it does not imply
	the absence of entanglement, i.e., we have no information if the 
	state is separable or entangled.
	
	In this work, we choose the Heisenberg Hamiltonian with nearest-neighbour 
	interaction,
	\begin{align}
		\label{eq:heisenberghamiltonian}
		H=-J\sum_{\langle i,j\rangle}(\sigma_{x}^{(i)}\sigma_{x}^{(j)}+\sigma_{y}^{(i)}\sigma_{y}^{(j)}+\sigma_{z}^{(i)}\sigma_{z}^{(j)})+h\sum_i\sigma_{z}^{(i)},
	\end{align}
	where $J$ is the coupling constant, $h$ is a external field and 
	$\sigma_{i}$ the Pauli matrices.
	This Hamiltonian was selected due to its wide applicability in diverse many-body systems as in 
	quantum batteries \cite{le2018spin,medina2023vqeinspired}, describing Kagome lattices \cite{bosse2022probing,kattemolle2022variational}, spin chains \cite{yamamoto1998elementary} and many-body localization \cite{vznidarivc2008many}.
	
	\section{Variational Quantum Eigensolver}\label{Sec.VQE}
	As mentioned before the VQAs are task-oriented algorithms aimed to minimize
	a cost function that encodes the solution of the problem \cite{cerezo2021cost}.
	Usually, the cost function consists in the expectation value of an
	observable $O$, which codifies the problem of interest,
	\begin{align}
		\label{costfunctiondef}
		\mathcal{C}(\boldsymbol{\theta})=\Tr\{OU(\boldsymbol{\theta})
		\rho_{0}\phantom{.}U^{\dagger}(\boldsymbol{\theta})\},
	\end{align}
	where $\rho_{0}$ represents an input state.
	The cost function describes a hyper-surface in terms of the parameters $\boldsymbol{\theta}$
	and the optimization process will seek the global minimum of this cost
	landscape. The optimal parameters $\boldsymbol{\theta}_{opt}$ represent 
	the optimal solution codified in the observable $O$ for a given Ansatz $U(\boldsymbol{\theta}_{opt})$,
	\begin{align}
		\label{eq:cost}
		\boldsymbol{\theta}_{opt}&=\stackrel[\boldsymbol{\theta}]{}{\mathrm{\phantom.arg\phantom{.} min}} \Tr\{OU(\boldsymbol{\theta})\rho_{0\phantom{.}}U^{\dagger}(\boldsymbol{\theta})\},
	\end{align}
	which is expected to obtain a result close enough to the real solution.
	
	This general framework shows the power of hybrid quantum-classical algorithms of
	this kind. A lot of problems that can be codified in a cost function of this type
	are potential candidates to be solved with VQAs. The cost function can be more general than the one posed
	in Eq. (\ref{costfunctiondef}), but must at least be measurable in a quantum
	device. Some examples are the use of geometrical distance between states \cite{Learning_qearthmover_cost} or operators \cite{costfunction_tracedistance_operators}.
	
	For our purposes, we chose the observable $O$, in Eq. (\ref{eq:cost}), as the Hamiltonian (\ref{eq:heisenberghamiltonian}). Then, our cost function represents the average value of the Heisenberg Hamiltonian parameterized by the Ansatz $U(\theta)$.
	This case is a specific type of VQA named Variational Quantum Eigensolver (VQE)
	\cite{benchmarking_eigensolvers,bosse2022probing,Mcclean_nature_VQE_photonic,tilly2022variational,cerezo2022stateeigensolver}.
	The cost function will be the mean energy value 
	\begin{align}\label{eq:vqeCost}
		\mathcal{C}(\boldsymbol{\theta}) = \Tr\{HU(\boldsymbol{\theta})\rho_{0}U^{\dagger}(\boldsymbol{\theta})\},
	\end{align}
	where $\rho_{0}$ is the initial state.
	The main interest of the VQE is to find the minimum value of the cost function, that 
	corresponds to the ground state energy. Once the cost function reaches the minimum, the optimized parameters, $ \boldsymbol{\theta}\to\boldsymbol{\theta}_{opt} $, 
	can be used to prepare the Hamiltonian ground state $\rho_{g}$,
	\begin{align}
		U(\boldsymbol{\theta}\to\boldsymbol{\theta}_{opt})\rho_{0}\phantom{.}U^{\dagger}(\boldsymbol{\theta}\to\boldsymbol{\theta}_{opt})\to\rho_{g}.
	\end{align}
	
	\section{Ansätze choice}\label{Sec.Ansatze}
	The parameterized quantum circuit used in a VQA is the \textit{Ansatz}.
	The Ansätze choice is an open problem in the proposal of methods
	and there are many consequences to the different options. Usually, this choice
	is heuristic and can either have motivation in the target problem or in the
	available hardware. 
	A great diversity of Ansätze were proposed in literature to either general \cite{HEA_proposal, dallaire2020application, cerezo2021variational} or specific tasks \cite{symmetryadapted_ansatz, buildingspaticalsymmetries_ansatz}, motivating works testing the performance of each one of them in the implementation of VQAs \cite{benchmarking_eigensolvers, benchmarking_quantumoptimalcontrol}.
	The selection of an Ansatz is therefore heuristic, so its choice may be motivated
	by the problem based on symmetries, the Hamiltonian of interest or entanglement \cite{choquette2021quantum,HVA_refprincipal,HEA_proposal}.
	
	Another reason to choose a specific Ansatz is the availability of the
	experimental platform and/or architecture of the quantum
	processor.
	In this section, we present the Ansätze that will be
	used in this work.
	To use the mean value of the entanglement witness (\ref{eq:ent.witness})
	as cost function, we need two different types of Ansätze: one to search for the lowest separable
	energy $E_{sep}$ and other to search for energy values lower than $E_{sep}$.
	This will configure the existence of an entanglement gap, i.e., the Hamiltonian
	have an entangled ground state.
	
	One of the most used Ansätze is the Hardware Efficient Ansatz (HEA), which employ
	the repetition of rotation layers and entangling gates that
	could connect all the qubits based on the architecture of the available 
	hardware.
	The main idea of this Ansatz is to minimize the circuit depth. The HEA 
	is also used when there is a hardware limitation due to the experimental
	platform available, such as the qubit connectivity.
	
\begin{figure}[!htb]
		\centering
		\includegraphics[width=.6\columnwidth]{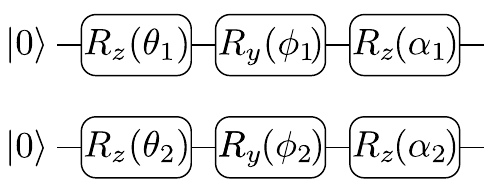}
		\caption{Hardware efficient separable Ansatz (HESA) for 2 qubits with only local rotations.}
		\label{Fig:HEA-separavel}
	\end{figure}

	To find the lowest separable energy (see Sec.\ref{Sec.EW}), we propose a Hardware Efficient Separable Ansatz (HESA)
	where only local rotations, $\left(R_{j}^{(\kappa)}(\theta_\ell)=e^{-i\theta_\ell\sigma^{(\kappa)}_{j}/2}\right)$, are performed in the qubits.
	the index $j=\{x,y,z\}$ denotes the axis of the rotation and $\kappa$ the qubit in which the rotation is applied.
	The rotations selected for this Ansatz, Fig. \ref{Fig:HEA-separavel}, consists in the 
	decomposition of a general unitary acting in one qubit \cite{nielsenchuangbook}.
	For a given initial separable
	state, the optimization process 
	searches for the lowest energy separable eigenstate of the Hamiltonian in Eq. (\ref{eq:heisenberghamiltonian}).
	The HESA is depicted in Fig. \ref{Fig:HEA-separavel}, where the
	initial state is fixed $\ket{\psi_0}=\ket{0}^{\otimes n}$, being $n$ is the 
	number of qubits.
	
	This first VQE process searches for the lowest separable energy, $E_{sep}$,
	to compose the energy-based entanglement witness.
	Scaling this Ansatz to more qubits will increase the number of trainable
	parameters in the same scale.
	
	\subsection{Hardware efficient Ansatz (HEA)}\label{Sec:HEA}
\begin{figure}[!htb]
		\centering
		\includegraphics[width=.8\columnwidth]{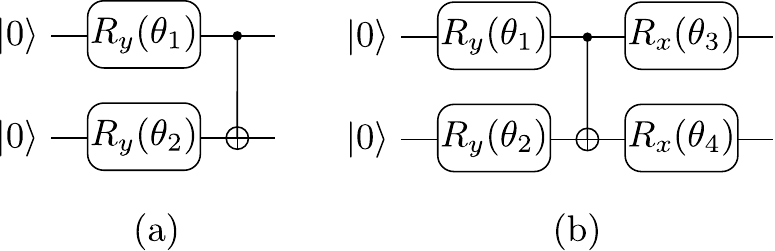}
		\caption{Hardware efficient Ansätze (HEA) for 2 qubits (a) one rotation and one CNOT layer HEA(a); (b) two rotation layers and one CNOT layer HEA(b).}
		\label{Fig:HEA}
	\end{figure}

	The class of hardware efficient Ansatz (HEA) is a broad set of circuits
	whose structure is problem-agnostic (its structure is not inspired on the studied
	problem) and its focus is to reduce the circuit depth.
	We selected two HEA circuits as presented in Fig. \ref{Fig:HEA}, 
	based on single-qubit rotations and CNOT layers.
	
	Increasing the number of qubits, the Ansatz structure remains as a single rotation 
	layer in each qubit, the CNOT layer applied sequentially in each pair
	of first neighbour qubits. The number of CNOT gates $\Lambda_{i,i+1}$ ($i= \mathrm{control}$, $i+1= \mathrm{target}$) in the Ansätze depicted in the Fig. \ref{Fig:HEA} is $n-1$, where $n$ is the number of qubits,
	\begin{align}
		U(\boldsymbol{\theta}) = \prod_{i=1}^{n} R_y^{(i)}(\theta_{i})
		\prod_{j=1}^{n-1}\Lambda_{j,j+1}, \label{eq:HEA(a)}
	\end{align}
	for the HEA(a) and
	\begin{align}
		U(\boldsymbol{\theta}) = \prod_{i=1}^{n} R_x^{(i)}(\theta_{i})
		\prod_{j=1}^{n-1}\Lambda_{j,j+1}
		\prod_{k=1}^{n}R_y^{(k)}(\theta_{k}), \label{eq:HEA(b)}
	\end{align}
	for the HEA(b), where the difference between these Ansätze is the addition of 
	a rotation layer $R_x^{(\cdot)}(\cdot)$.
	In both cases of Fig. \ref{Fig:HEA} we tried the simplest possible rotations,
	where the $R_y(\theta_i)$, Fig.\ref{Fig:HEA} (a), presented the best results. 
	Using different combinations of local rotations in Fig. \ref{Fig:HEA} (b), we did not see a significant discrepancy among the results.
	
	\subsection{Sycamore-inspired Ansatz}
\begin{figure}[!htb]
		\centering
		\includegraphics[width=.7\columnwidth]{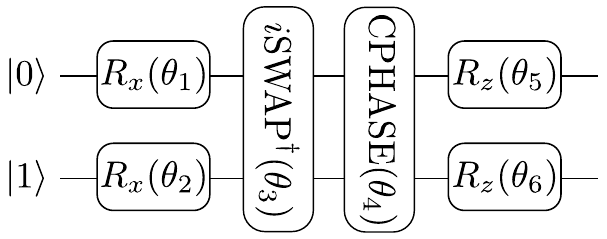}
		\caption{The Sycamore-inspired Ansatz for two qubits.}
		\label{Fig:Syc}
	\end{figure}

	Another selected Ansatz to the benchmark is inspired by Sycamore
	Google's chip \cite{arute2019quantum,katabarwa2022connecting}.
	Its structure is described in \cite{dallaire2020application} and is 
	depicted in the Fig. \ref{Fig:Syc}.
	This Ansatz may be classified as HEA because its structure is fixed and
	problem agnostic. This Ansatz is built using local rotation gates,
	iSWAP and CPHASE for two qubits operations given by
	\begin{align}
		i\mathrm{SWAP}^{\dagger}(\theta)=\exp\left[-i\theta\left(\sigma^{(1)}_{x}\sigma^{(2)}_{x}+\sigma^{(1)}_{y}\sigma^{(2)}_{y}\right)\right],
	\end{align}
	\begin{align}
		\mathrm{CPHASE}(\phi)=e^{-i\phi \sigma^{(1)}_{z} \sigma^{(2)}_{z}},
	\end{align}
	and in this Ansatz, the initial state is prepared as an alternated product of
	$\ket{0}$ and $\ket{1}$.

	It is important to remark that the HEA have the drawback of not
	being efficient to some problems due to the lack of problem inspiration \cite{wu2021efficient}
	and to the generation of great difficulty in training the cost function \cite{cerezo2021cost}.
	
	\subsection{Hamiltonian Variational Ansatz (HVA)}
	
	The Hamiltonian Variational Ansatz (HVA) is inspired by one of the most famous VQA: the Quantum Alternating Operator Ansatz (QAOA) \cite{farhi2014qaoa,zhou2020quantum}, used mainly for combinatorial optimization problems.
	The QAOA consists in the alternated application of the circuit digitalization of the evolution of two non-commuting Hamiltonians: a Hamiltonian $H_P$ in terms of products of the Pauli-$z$ matrices which encodes the problem of interest, and a Hamiltonian $H_x$ in terms of products of Pauli-$x$ matrices, to introduce coherences in the $z$-basis. This alternated application emulates the evolution of the total Hamiltonian $H=H_P + H_x$ via the Trotter-Suzuki decomposition \cite{nielsenchuangbook}
	\begin{align}
		e^{H_P+H_x}=\lim_{n\to\infty}(e^{H_P /n}e^{H_x /n})^n,
	\end{align}
	therefore, the unitary evolution of the system becomes
	\begin{align}
		U=\lim_{n\rightarrow\infty}\prod^n_{k=1}e^{-iH_P t/n}e^{-iH_x t/n}.
	\end{align}
	The QAOA Ansatz is implemented by substituting the terms $t/n$ with circuit parameters and taking the number of applications $n$ to be of finite size instead, making the experimental implementation possible. The Ansatz is an approximation of the Trotter-Suzuki decomposition but it is still reliable. In fact, this restriction does not affect the possibility of building the desired state by the optimization of the parameters.
	
	If, instead of two non-commuting terms, the Hamiltonian is decomposed in a bigger number of non-commuting terms, like in the Heisenberg Hamiltonian, $H=\sum_jH_j$, the Trotter-Suzuki decomposition reads
	\begin{align}
		U\approx\prod_{k=1}^n\prod_je^{-iH_jt/n},
	\end{align}
	which is the decomposition for the HVA Ansatz. Each $H_j$ represents one of the noncommuting terms.
	
\begin{figure}[!htb]
		\centering
		\includegraphics[width=\columnwidth]{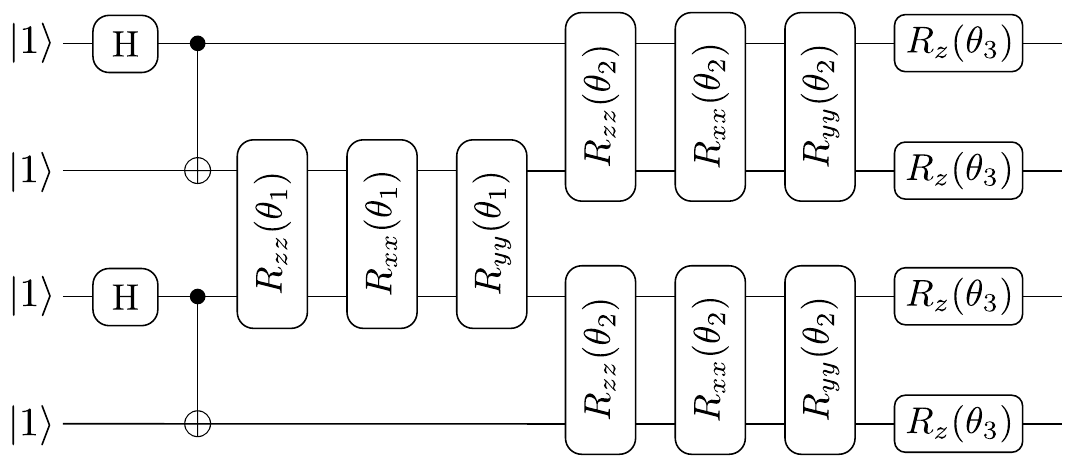}
		\caption{Structure of Hamiltonian Variational Ansatz (HVA) for 4 qbits.}
		\label{Fig:HVA}
	\end{figure}

	Inspired by Refs. \cite{HVA_refprincipal, HVA_refadicional}, we split the couplings between qubits into even and odd links, thus each term will be of the form $H_{\kappa\kappa}=H_{\kappa\kappa}^{\mathrm{odd}}+H_{\kappa\kappa}^{\mathrm{even}}$. Exploiting the Heisenberg Hamiltonian symmetry, which has the same coupling constant $J$ for each direction, we chose the parameterization such that the even links, odd links and individual terms of the external field have the same angles in each set. This decomposition for only one layer of the circuit will be in the form (omitting the state preparation)
	\begin{widetext}
		\begin{align}
			U(\theta_1,\theta_2,\theta_3)\approx     
			G(\theta_3,H_z)G(\theta_2,H_{yy}^{\mathrm{even}})G(\theta_2,H_{xx}^{\mathrm{even}})G(\theta_2,H_{zz}^{\mathrm{even}})G(\theta_1,H_{yy}^{\mathrm{odd}})G(\theta_1,H_{xx}^{\mathrm{odd}})G(\theta_1,H_{zz}^{\mathrm{odd}}),   
		\end{align}
	\end{widetext}
	where $G(\theta,H):=e^{-i\theta H}$. The initial state is the ground state of the sum of the even parts of the total Hamiltonian, i.e.,
	\begin{align}
		\ket{\psi_0}=\bigotimes_{i=1}^{N/2}\frac{1}{\sqrt{2}}\left(\ket{01}-\ket{10}\right)_{2i-1,2i}.
		\label{Eq:InitialStateHVA}
	\end{align}
	This is why the execution of the unitary is initialized with the odd part, so the initial state is not an eigenstate. Fig. \ref{Fig:HVA} presents the state preparation and execution of HVA for a system with 4 qubits.
	
	The gates applied to one or two qubits are of the form
	\begin{align}
		R_{\alpha}^{(j)}(\theta_\ell)=e^{-i \theta_\ell \alpha^{(j)}/2},
	\end{align}
	where $\alpha$ is the Pauli strings $\left(x, y, z, xx, xy,...\right)$
	corresponding to the Pauli matrices and its products 
	$\left(\sigma_{x}, \sigma_{y}, \sigma_{z}, \sigma_{x}\otimes\sigma_{x}, \sigma_{x}\otimes\sigma_{y},...\right)$ applied to the $j$ qubit or pairs of qubits.
	For odd number of qubits, the remaining \aspasc not paired\aspasd 
	qubit is connected with the others just by the first rotation layer.
	
	\subsection{Low-Depth Circuit Ansatz (LDCA)}
\begin{figure}[!htb]
		\centering
		\includegraphics[width=0.8\columnwidth]{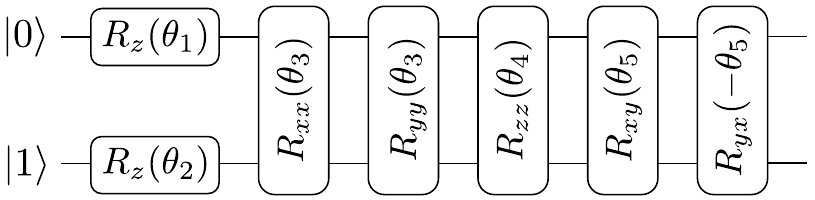}
		\caption{The Low-Depth Circuit Ansatz (LDCA) for two qubits.}
		\label{Fig:LDCA}
	\end{figure}

	The last selected Ansatz that we choose to benchmark is the low-depth 
	circuit Ansatz presented in Ref. \cite{katabarwa2022connecting}.
	In Fig. \ref{Fig:LDCA} we depict this Ansatz for two qubits. 
	
	This Ansatz was proposed in Ref. \cite{dallaire2019low} with a similar
	structure where the Ansatz of Fig. \ref{Fig:LDCA} is a part of the 
	decomposition of the complete LDCA.
	
	Scaling the number of qubits, these operations are applied to all
	pairs of adjacent qubits (omitting the circuit $\boldsymbol{\theta}$ parameters),
	\begin{align}
		U(\boldsymbol{\theta})=\prod_{i=1}^{n-1}R^{(i,j)}_{yx}R^{(i,j)}_{xy}
		R^{(i,j)}_{zz}R^{(i,j)}_{yy}R^{(i,j)}_{xx}
		R^{(j)}_{z}R^{(i)}_{z},
	\end{align}
	where $j=i+1$.
	
	\section{Results}\label{Sec.Results}
	In this section, we present the procedure and results from the Ansätze benchmark.
	As described in Sec. \ref{Sec.Ansatze} we need two Ansätze to perform the entanglement
	detection using the mean value of witness Eq. (\ref{eq:ent.witness}) as a cost function.
	For now on, we will take the coupling $J=-1$ and the absence of the external
	field $h=0$ for all the simulations.
	Each Ansatz has a specific initial state, as depicted in each respective figure.
	For all the simulations we use the Pennylane library \cite{bergholm2018pennylane}.
	
	The first analysis was about the number of shots, then the entanglement detection, and 
	finally the ground state convergence, varying the number of qubits. Each analysis considered all chosen Ansätze.
	
	\subsection{Number of Shots}
	The number of shots in a variational algorithm impacts cost function estimation.
	In an experimental context, the quantum hardware will run the circuit and perform the measurements in the $\sigma_z$ basis in each wire of the circuit, i.e., in each qubit. This procedure, called shot, is repeated many times in order to obtain the relevant statistics of a certain observable, such as its mean value and variance.  In the idealized scenario, an infinity number of measurements would be required to obtain the exactly observable quantities. Such infinity number of measurements is, however, impossible to do in practice.
	
	To understand how the number of shots impact the cost function estimation, we ran our VQE simulations considering 10, 50, 100, and 300 shots. As the VQE simulation takes random initial parameters, and the initial conditions can strongly influence in the cost function convergence, we ran the VQE protocol 250 times for each fixed number of shots, each time taking a different set of initial random parameters. As an example, for the case with 10 shots, we proceed as follows. First, a random set of parameters is selected in the interval $[0,2\pi]$. With this set, we collect 10 shots. Then, we proceed taking 10 shots in each iteration step. We verified that 200 interactions are good enough to make the cost function converge. After the convergence, we repeat these steps for another set of arbitrary initial parameters. We repeat this process 250 times, and finally compute the average cost function convergence. Moreover, we compare the average cost function using different number of shots with the average cost function obtained from the ideal case, i.e., the cost function analytically computed assuming that we have the complete information about the system’s state, which is given by $U(\boldsymbol{\theta})\rho_0U^\dagger(\boldsymbol{\theta})$.
	
	\begin{figure*}[!htb]
		\centering
		\includegraphics[width=2\columnwidth]{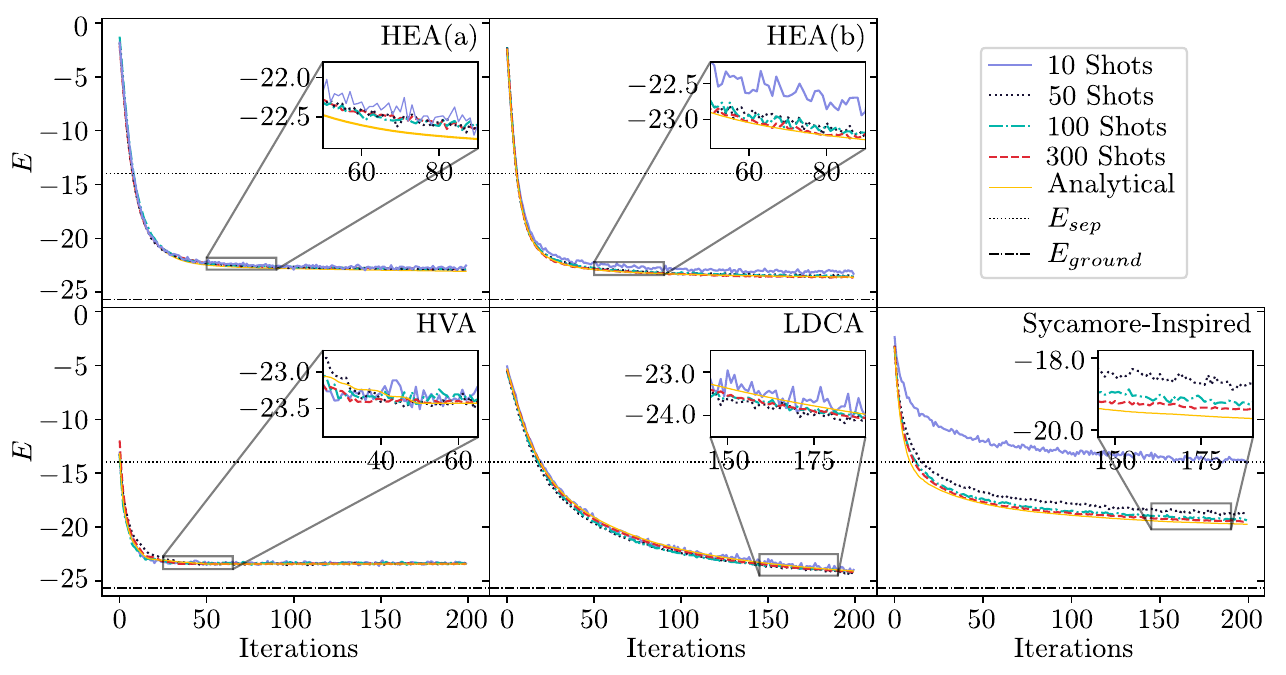}
		\caption{(Colour online) Optimization trajectory of ground state convergence for
			the HEA(a), HEA(b), HVA, LDCA and Sycamore-inspired Ansätze. These curves are the mean value
			over 250 random starts with 15 qubits. The zoom was applied to
			visualize the proximity among different numbers of shots in each case.  For 15 qubits $E_{sep}=-14$ and $E_{ground}=-25.6$}
		\label{Fig:shots}
	\end{figure*}
	
	As presented in Fig. \ref{Fig:shots}, we set a limit of 200 iterative steps 
	to the VQE optimization. Most of the Ansätze (HEA(a), HEA(b), HVA) achieves a minimum value of 100 steps. The simulated VQE does not achieve the global minimum in these Ansätze. 
	The LDCA will achieve the minimum in fewer steps, whereas Sycamore-inspired needs even more steps to minimize the cost function.
	The Fig. \ref{Fig:shots} presents the mean value over 250 runs of each VQE. There are cases where the cost function reaches the global minimum (ground-state energy). On the other side, there are runs where the optimization gets imprisoned in a local minimum, leading the mean value to not achieve the global minimum.
	
	The influence of shots in the Ansätze HEA(a), HVA, and LDCA presents 
	fluctuations close to the analytic calculation and the results are very close
	for all the shots presented in the Fig. \ref{Fig:shots}.
	In the HEA(b) with 10 shots, there is a discrepancy in the mean value,
	where the optimization is worse than the other number of shots.
	The most affected case is for the Sycamore-inspired with 10 shots, where the 
	entanglement witnessing criterion (Sec.\ref{Sec:Entanglement Detection}) would occur after the 200 iterative steps.
	Increasing the number of shots, the optimization achieves better results, becoming 
	closer to the analytic case.
	
	As the optimization trajectory of the cost function has a similar behaviour to the analytic
	calculation (as we might see looking at the zoom in each Ansatz) for most of the presented
	cases, we chose to work with 100 shots.
	
	\subsection{Entanglement Detection}\label{Sec:Entanglement Detection}
\begin{figure}[!htb]
		\centering
		\includegraphics[width=\columnwidth]{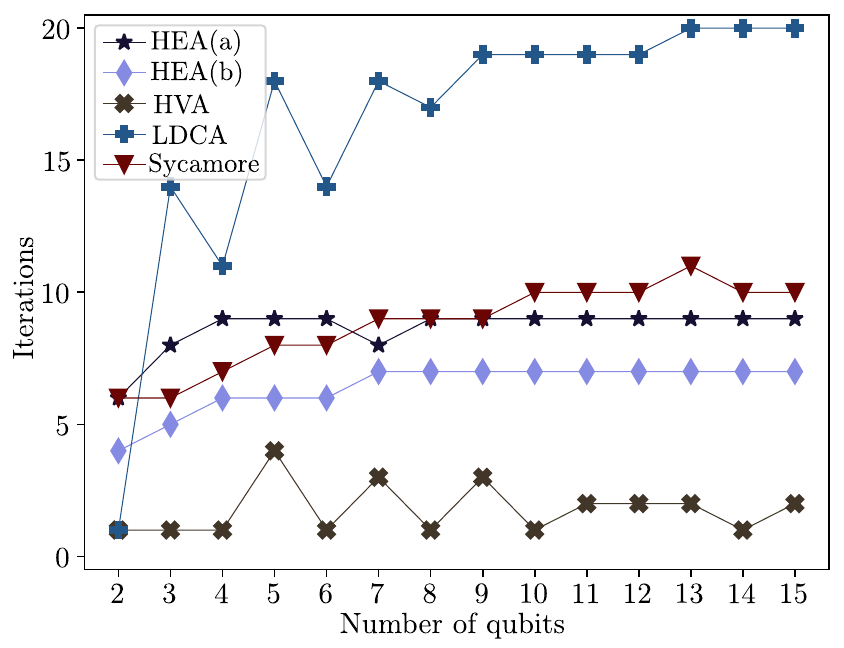}
		\caption{(Colour online) Iterations needed to detect entanglement
			using the energy-based entanglement witness for 2 to 15
			qubits for each Ansatz. }
		\label{Fig:EW}
	\end{figure}

	The entanglement detection: we run the VQE starting from the fixed state
	$\ket{\psi_0}=\ket{0}^{\otimes n}$
	and optimize the parameters of HESA to search for the minimum separable energy $E_{sep}$.
	Using the other Ansätze with entangling gates, we run a second VQE to seek for energy
	values lower than $E_{sep}$, i.e., to detect entanglement using the witness presented in
	Eq. (\ref{eq:ent.witness}).
	In Fig. \ref{Fig:EW} we present the number of optimization steps needed to detect
	entanglement using the selected Ansatz from 2 to 15 qubits.
	As the initial set of parameters are randomly chosen (in the interval $\left[0,2\pi\right]$),
	we run the optimization process
	250 times to compute the mean number of steps needed to detect entanglement.
	
	We note in Fig. \ref{Fig:EW} that the entanglement detection occurs in a few
	optimization steps, where the LDCA needs more steps to detect entanglement
	compared to the other Ansätze.
	In HEA(a) the entanglement detection occurs in a few optimization steps and
	there is an improvement in HEA(b), where fewer
	steps are needed to detect entanglement. This is observed even when scaling the number
	of qubits.
	It is a very interesting case because both circuits 
	have the same entanglement capability \cite{correr2024characterizing}.
	The one that needs fewer steps is the HVA, due to the initial state being 
	prepared as presented in Eq. (\ref{Eq:InitialStateHVA}).
	Even for LDCA that demands more steps, all the Ansätze 
	present an entanglement detection in a few optimization steps.
	So instead of setting a stop condition at entanglement detection,
	we let the VQE free to run until 200 optimization steps, to search for
	the entangled ground state.
	
	\subsection{Benchmark}
\begin{figure}[!htb]
		\centering
		\includegraphics[width=\columnwidth]{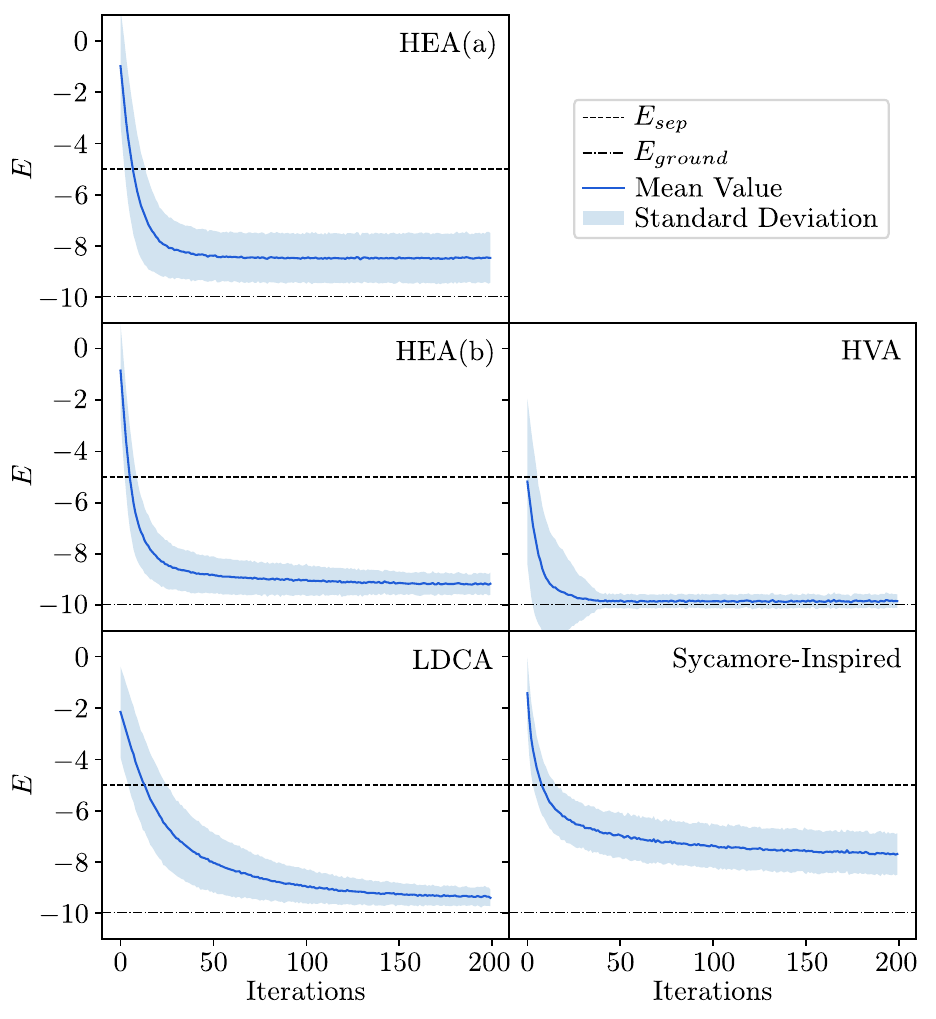}
		\caption{(Colour online) Ground state convergence and
			standard deviation over 250 random starts for the HEA(a) and (b),
			HVA, LDCA, and Sycamore-inspired Ansatz with 6 qubits. For 6 qubits we have $E_{sep}=-5$, and $E_{ground}=-9.9$. }
		\label{Fig:6qbitsExample}
	\end{figure}

	As mentioned before, we run 250 random starts for each Ansatz to 
	compute the mean value and the standard deviation.
	In Fig. \ref{Fig:6qbitsExample} we present an example of the
	procedure for 6 qubits. The same was done by scaling the number
	of qubits.
	
	\begin{figure}[!htb]
		\centering
		\includegraphics[width=\columnwidth]{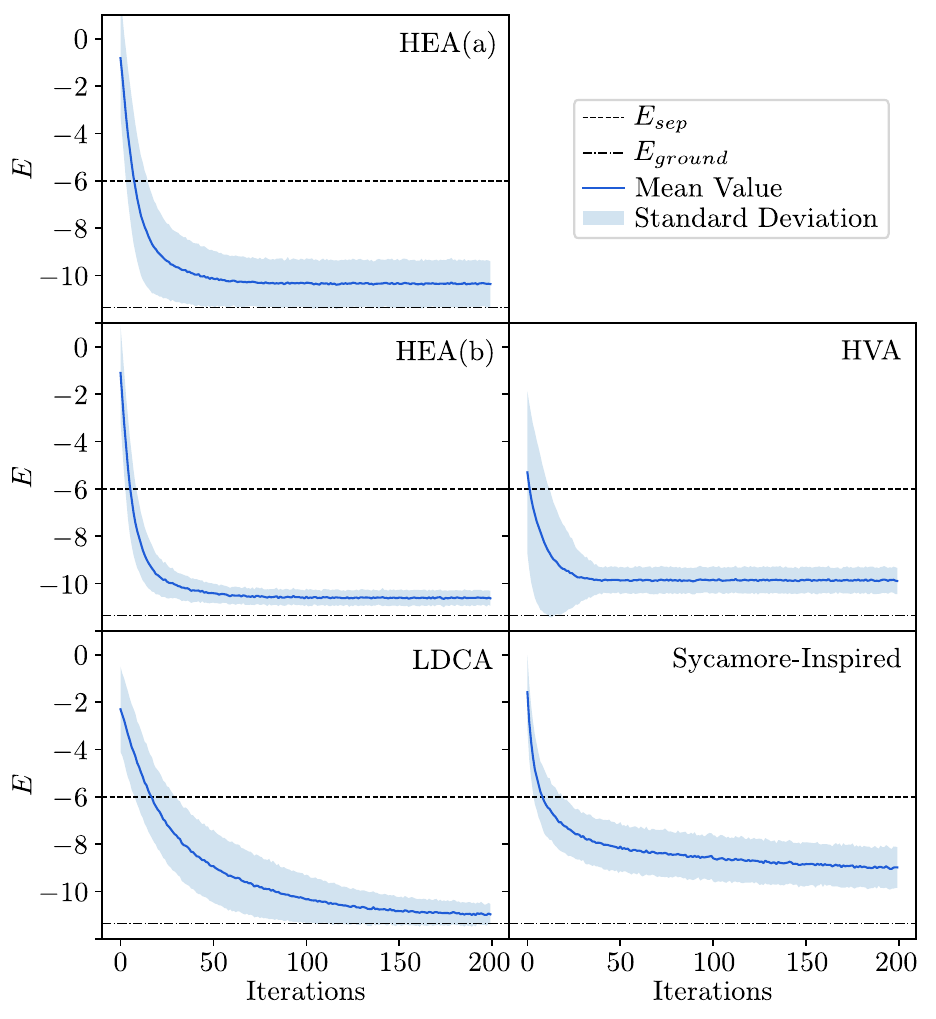}
		\caption{(Colour online) Ground state convergence and
			standard deviation over 250 random starts for the HEA(a) and (b),
			HVA, LDCA, and Sycamore-inspired Ansatz with 7 qubits. For 7 qubits we have $E_{sep}=-6$, and $E_{ground}=-11.3$.}
		\label{Fig:7qbitsExample}
	\end{figure}
	
	Based on the Fig. \ref{Fig:6qbitsExample}, 
	we note a large dispersion in standard deviation in HEA(a) that becomes narrow
	in HEA(b), this improvement occurs due to the additional rotation layer. Also, the
	mean value of the convergence gets closer to the ground state energy value.
	The best convergence to the entangled ground state is observed in HVA,
	followed by the LDCA.
	The Sycamore-inspired Ansatz needs more optimization steps to reach the ground state,
	spending a great computational time as the number of qubits 
	increases.
	
	In the HEAs, the convergence is better for an odd number of qubits as
	presented in Fig. \ref{Fig:7qbitsExample},
	while the HVA has better convergence with an even number of qubits.
	The LDCA and Sycamore-inspired presented very similar behaviour in both cases.
	
	\section{Discussion and outlook}\label{Sec.Discuss}
	Initially, our goal was to use the VQE with different Ansätze to
	witness ground state entanglement by using the energy as a part of the witness,
	where the Heisenberg Hamiltonian describes the many-body system.
	However, as presented in Sec. \ref{Sec:Entanglement Detection},
	the witnessing occurs in a few optimization steps.
	This is an important result, once this witnessing process could be used
	to verify the existence of an entangled ground state in more complex Hamiltonians.
	As described in Sec. \ref{Sec.EW} \cite{dowling2004energy},
	if there is a state with an energy value lower than the minimum 
	separable, it implies the existence of an entangled ground state.
	So, instead of stopping the VQEs when the witnessing occurs, we search
	for the entangled ground state running the VQEs until 200 iterative
	steps.
	
	After running for multiple qubits and shots choices, we decided
	to work with 100 shots. In Fig. \ref{Fig:shots} we present for 15 qubits these results as an
	example of this choice.
	
	The Fig. \ref{Fig:EW} presents the number of iterations needed to
	witness entanglement, using the Eq. (\ref{eq:ent.witness}) for all
	the selected Ansätze where the number of qubits varies from 2 to 15.
	The many-body system is modelled with the Heisenberg Hamiltonian in the absence of external field (Eq. \ref{eq:heisenberghamiltonian}).
	
\begin{figure}[!htb]
		\centering
		\includegraphics[width=\columnwidth]{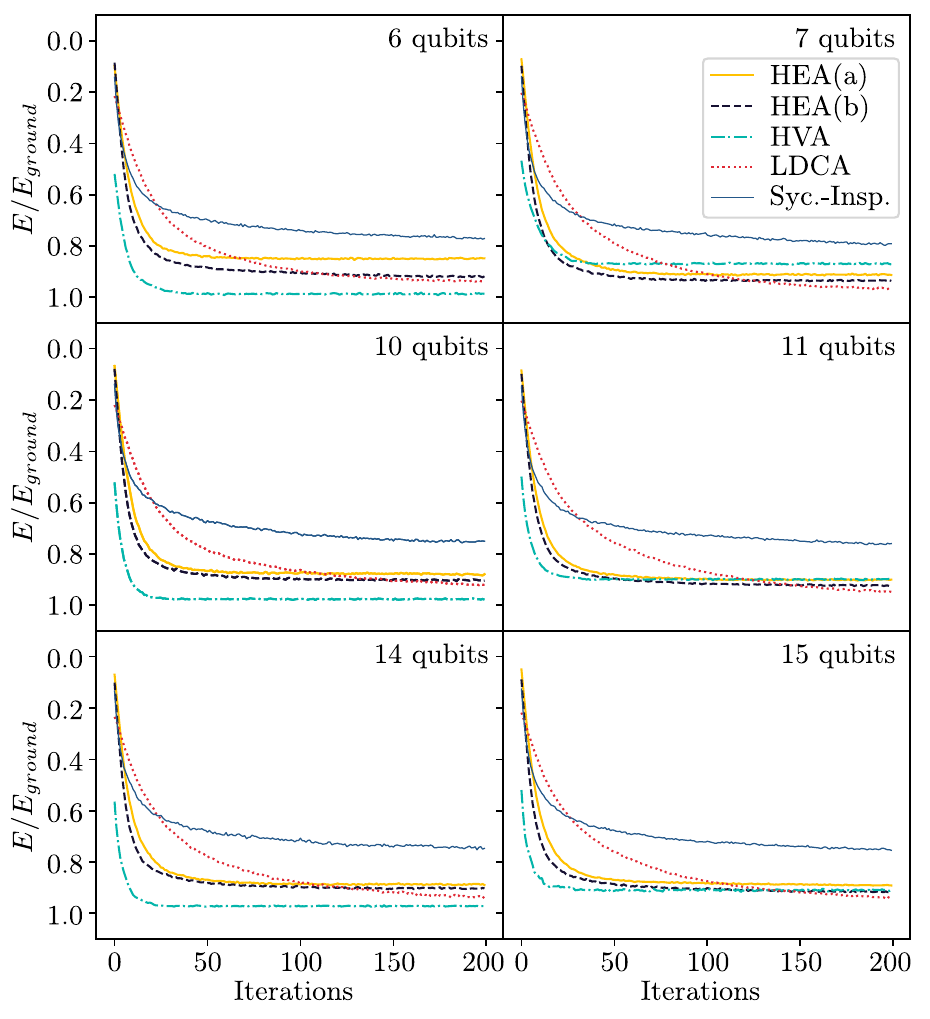}
		\caption{(Colour online) Mean value of ground state
			convergence normalized by the calculated ground state energy for
			all selected Ansätze and different numbers of qubits.}
		\label{Fig:NormalizedConvergence}
	\end{figure}

	In Fig. \ref{Fig:NormalizedConvergence} we present the ground
	state convergence normalized by the calculated ground state energy\footnote{The \textit{calculated value of ground state energy} was found solving the Hamiltonian numerically.} for each 
	Ansätze and number of qubits presented.
	
	As the number of qubits increases, the number of trainable parameters
	also increases. This could lead to some optimization problems increasing
	the computational time or/and barren plateaus \cite{cerezo2021variational}, which are regions in
	the cost landscape where the gradient vanishes, not giving a preferred
	path to the optimization process. In Table \ref{Table:parameters} we
	present the scale of trainable parameters associated with the number
	of qubits used in each Ansatz.
	
	\begin{table}[!htb]
		\centering
		\begin{tabular}{l|c}
			\hline
			{\bf \ Ansatz} & {\bf \ Number of parameters\ }\\
			\hline\hline
			\ HEA(a) & \  $n_q$  \\ \hline
			\ HEA(b) & \  $2 n_q$    \\ \hline
			\ HVA & \  $3$    \\ \hline
			\ LDCA & \  $5 n_q$    \\ \hline
			\ Sycamore-inspired \ & \ $6n_q$ \  \\ \hline 
		\end{tabular}
		\caption{Scale of trainable parameters associated to the number of qubits $n_q$ for each Ansatz in a VQE process.} \label{Table:parameters}
	\end{table}
	
	Looking at the ground state convergence, we show the mean value and
	standard deviation for 6 and 7 qubits (Figs. \ref{Fig:6qbitsExample}
	and \ref{Fig:7qbitsExample}). The HEA(a) presented a large dispersion
	in standard deviation. This possibly is associated with the concentration
	of local minima, observed in some optimization trajectories.
	Modifying the HEA(a) with an extra
	rotation layer we have HEA(b), where the optimization is
	improved. The dispersion
	is reduced and the mean value got closer to the ground state as we can
	see in Fig. \ref{Fig:NormalizedConvergence}. This improvement is
	observable for even and also odd numbers of qubits.
	
	The HVA is an interesting case of a problem-inspired Ansatz where 
	the gates are similar to the Hamiltonian interactions and before
	the parameterized quantum circuit there is a state preparation 
	step. It was proposed in Ref. \cite{HVA_refprincipal}, where this 
	Ansatz construction is to work with an even number of qubits, so the worst
	efficiency with odd qubits was expected, due to the last qubit
	not being connected to the other ones.
	This is the Ansatz that
	presents the best convergence (for an even number of qubits) in a few
	optimization steps and very narrow standard deviation.
	One interesting point about this Ansatz is the 
	number of trainable parameters, that remain unchanged as the number 
	of qubits is increased.
	
	The LDCA presented a good performance to reach the ground state. 
	However, by scaling the number of qubits, more iterative steps are needed
	to reach the ground state as we can see in Figs.
	\ref{Fig:6qbitsExample}, \ref{Fig:7qbitsExample} and
	\ref{Fig:NormalizedConvergence}, where the standard deviation
	becomes narrower around 200 iterations. As the number of parameters
	increases with a factor of 5 of qubits, this Ansatz spends a great
	computational time to run.
	
	Finally, the Sycamore-inspired Ansatz is the one we obtained the
	worst performance among the selected Ansätze. This is the Ansatz that
	spends the greatest number of optimization steps to obtain
	the ground state. This could be assigned to the scale of parameters
	that increase with a factor 6 of the number of qubits. The parameters scale,
	in certain cases, might improve the optimization task, for example increasing 
	the number of layers. For our purposes with only one layer, this Ansatz
	is the one that needs more iterative steps to achieve a minimum.
	
	It is important to remark that the VQAs structure consists of an Ansatz,
	cost function, and optimization process. In this paper we fixed the
	cost function and optimization method, just working with Ansätze
	changes. So the efficiency of each case could be changed by working
	with other VQA characteristics to obtain the best match of these
	aspects.
	
	\begin{acknowledgments}
		\noindent This study was financed in part by the Coordenação de Aperfeiçoamento de Pessoal de Nível Superior – Brasil (CAPES) – Finance Code 001 (A.D. and G.I.C.).
		I.M. acknowledges financial support from São Paulo Research Foundation - FAPESP (Grants No. 2022/08786-2 and No. 2023/14488-7).
		P.C.A. acknowledges financial support from Conselho Nacional de Desenvolvimento
		Científico e Tecnológico - Brazil (CNPq - Grant No. 160851/2021-1).
		A.C. acknowledges a license by the Federal University of Alagoas for a sabbatical at the University of São Paulo, and partial financial support by CNPq (Grant No. 168785/2023-4), Alagoas State Research Agency (FAPEAL) (Grant No. APQ2022021000153), and São Paulo Research Foundation (FAPESP) (Grant No. 2023/03562-1).
		D.O.S.P. acknowledges the support by the Brazilian funding agencies CNPq (Grants No.
		304891/2022-3 and No. 402074/2023-8), FAPESP (Grant No. 2017/03727-0) and the Brazilian National
		Institute of Science and Technology of Quantum Information (INCT/IQ).
	\end{acknowledgments}
	

\begin{thebibliography}{10}
	
	\bibitem{preskill2018quantum}
	John Preskill.
	\newblock Quantum computing in the nisq era and beyond.
	\newblock {\em Quantum}, 2:79, 2018.
	
	\bibitem{leymann2020bitter}
	Frank Leymann and Johanna Barzen.
	\newblock The bitter truth about gate-based quantum algorithms in the nisq era.
	\newblock {\em Quantum Science and Technology}, 5(4):044007, 2020.
	
	\bibitem{RevModPhys.94.015004}
	Kishor Bharti, Alba Cervera-Lierta, Thi~Ha Kyaw, Tobias Haug, Sumner
	Alperin-Lea, Abhinav Anand, Matthias Degroote, Hermanni Heimonen, Jakob~S.
	Kottmann, Tim Menke, Wai-Keong Mok, Sukin Sim, Leong-Chuan Kwek, and Al\'an
	Aspuru-Guzik.
	\newblock Noisy intermediate-scale quantum algorithms.
	\newblock {\em Rev. Mod. Phys.}, 94:015004, Feb 2022.
	
	\bibitem{bharti2022noisy}
	Kishor Bharti, Alba Cervera-Lierta, Thi~Ha Kyaw, Tobias Haug, Sumner
	Alperin-Lea, Abhinav Anand, Matthias Degroote, Hermanni Heimonen, Jakob~S
	Kottmann, Tim Menke, et~al.
	\newblock Noisy intermediate-scale quantum algorithms.
	\newblock {\em Reviews of Modern Physics}, 94(1):015004, 2022.
	
	\bibitem{callison2022hybridalgorithms}
	Adam Callison and Nicholas Chancellor.
	\newblock Hybrid quantum-classical algorithms in the noisy intermediate-scale
	quantum era and beyond.
	\newblock {\em Phys. Rev. A}, 106:010101, Jul 2022.
	
	\bibitem{cerezo2021variational}
	Marco Cerezo, Andrew Arrasmith, Ryan Babbush, Simon~C Benjamin, Suguru Endo,
	Keisuke Fujii, Jarrod~R McClean, Kosuke Mitarai, Xiao Yuan, Lukasz Cincio,
	et~al.
	\newblock Variational quantum algorithms.
	\newblock {\em Nature Reviews Physics}, 3(9):625--644, 2021.
	
	\bibitem{cerezo2021cost}
	Marco Cerezo, Akira Sone, Tyler Volkoff, Lukasz Cincio, and Patrick~J Coles.
	\newblock Cost function dependent barren plateaus in shallow parametrized
	quantum circuits.
	\newblock {\em Nature communications}, 12(1):1791, 2021.
	
	\bibitem{HEA_proposal}
	Abhinav Kandala, Antonio Mezzacapo, Kristan Temme, Maika Takita, Markus Brink,
	Jerry~M. Chow, and Jay~M. Gambetta.
	\newblock Hardware-efficient variational quantum eigensolver for small
	molecules and quantum magnets.
	\newblock {\em Nature}, 549(7671):242--246, September 2017.
	
	\bibitem{HVA_refprincipal}
	Roeland Wiersema, Cunlu Zhou, Yvette de~Sereville, Juan~Felipe Carrasquilla,
	Yong~Baek Kim, and Henry Yuen.
	\newblock Exploring entanglement and optimization within the hamiltonian
	variational ansatz.
	\newblock {\em PRX Quantum}, 1:020319, Dec 2020.
	
	\bibitem{McClean_Ucoupledcluster}
	Jarrod~R McClean, Jonathan Romero, Ryan Babbush, and Alán Aspuru-Guzik.
	\newblock The theory of variational hybrid quantum-classical algorithms.
	\newblock {\em New Journal of Physics}, 18(2):023023, feb 2016.
	
	\bibitem{bauer2020quantum}
	Bela Bauer, Sergey Bravyi, Mario Motta, and Garnet Kin-Lic Chan.
	\newblock Quantum algorithms for quantum chemistry and quantum materials
	science.
	\newblock {\em Chemical Reviews}, 120(22):12685--12717, 2020.
	
	\bibitem{delgado2021variational}
	Alain Delgado, Juan~Miguel Arrazola, Soran Jahangiri, Zeyue Niu, Josh Izaac,
	Chase Roberts, and Nathan Killoran.
	\newblock Variational quantum algorithm for molecular geometry optimization.
	\newblock {\em Physical Review A}, 104(5):052402, 2021.
	
	\bibitem{medina2023vqeinspired}
	Ivan Medina, Alexandre Drinko, Guilherme~I Correr, Pedro~C Azado, and Diogo~O
	Soares-Pinto.
	\newblock Variational-quantum-eigensolver--inspired optimization for spin-chain
	work extraction.
	\newblock {\em Physical Review A}, 110(1):012443, 2024.
	
	\bibitem{VQErg}
	Duc~Tuan Hoang, Friederike Metz, Andreas Thomasen, Tran~Duong Anh-Tai, Thomas
	Busch, and Thom\'as Fogarty.
	\newblock Variational quantum algorithm for ergotropy estimation in quantum
	many-body batteries.
	\newblock {\em Phys. Rev. Res.}, 6:013038, Jan 2024.
	
	\bibitem{verdon2019quantum}
	Guillaume Verdon, Jacob Marks, Sasha Nanda, Stefan Leichenauer, and Jack
	Hidary.
	\newblock Quantum hamiltonian-based models and the variational quantum
	thermalizer algorithm, 2019.
	
	\bibitem{consiglio2023variational}
	Mirko Consiglio, Jacopo Settino, Andrea Giordano, Carlo Mastroianni, Francesco
	Plastina, Salvatore Lorenzo, Sabrina Maniscalco, John Goold, and Tony~JG
	Apollaro.
	\newblock Variational gibbs state preparation on nisq devices.
	\newblock {\em arXiv preprint arXiv:2303.11276}, 2023.
	
	\bibitem{warren2022adaptive}
	Ada Warren, Linghua Zhu, Nicholas~J Mayhall, Edwin Barnes, and Sophia~E
	Economou.
	\newblock Adaptive variational algorithms for quantum gibbs state preparation.
	\newblock {\em arXiv preprint arXiv:2203.12757}, 2022.
	
	\bibitem{silva2024simulating}
	Ana~Clara das Neves~Silva and Clebson Cruz.
	\newblock Simulating thermodynamic properties of dinuclear metal complexes
	using variational quantum algorithms, 2024.
	
	\bibitem{uvarov2020machine}
	AV~Uvarov, AS~Kardashin, and Jacob~D Biamonte.
	\newblock Machine learning phase transitions with a quantum processor.
	\newblock {\em Physical Review A}, 102(1):012415, 2020.
	
	\bibitem{canabarro2022quantum}
	Askery Canabarro, Taysa~M. Mendonça, Ranieri Nery, George Moreno, Anton~S.
	Albino, Gleydson~F. de~Jesus, and Rafael Chaves.
	\newblock Quantum finance: um tutorial de computação quântica aplicada ao
	mercado financeiro.
	\newblock {\em Revista Brasileira de Ensino de Física}, 44:e20220099, 2022.
	
	\bibitem{wang2024variational}
	Shengbin Wang, Peng Wang, Guihui Li, Shubin Zhao, Dongyi Zhao, Jing Wang, Yuan
	Fang, Menghan Dou, Yongjian Gu, Yu-Chun Wu, et~al.
	\newblock Variational quantum eigensolver with linear depth problem-inspired
	ansatz for solving portfolio optimization in finance.
	\newblock {\em arXiv preprint arXiv:2403.04296}, 2024.
	
	\bibitem{brandhofer2022qaoaportfolio}
	Sebastian Brandhofer, Daniel Braun, Vanessa Dehn, Gerhard Hellstern, Matthias
	H\"{u}ls, Yanjun Ji, Ilia Polian, Amandeep~Singh Bhatia, and Thomas Wellens.
	\newblock Benchmarking the performance of portfolio optimization with {QAOA}.
	\newblock {\em Quantum Information Processing}, 22(1), December 2022.
	
	\bibitem{liu2023practical}
	Chen-Yu Liu.
	\newblock Practical quantum search by variational quantum eigensolver on noisy
	intermediate-scale quantum hardware.
	\newblock {\em arXiv preprint arXiv:2304.03747}, 2023.
	
	\bibitem{havlivcek2019supervised}
	Vojt{\v{e}}ch Havl{\'\i}{\v{c}}ek, Antonio~D C{\'o}rcoles, Kristan Temme,
	Aram~W Harrow, Abhinav Kandala, Jerry~M Chow, and Jay~M Gambetta.
	\newblock Supervised learning with quantum-enhanced feature spaces.
	\newblock {\em Nature}, 567(7747):209--212, 2019.
	
	\bibitem{biamonte2017quantum}
	Jacob Biamonte, Peter Wittek, Nicola Pancotti, Patrick Rebentrost, Nathan
	Wiebe, and Seth Lloyd.
	\newblock Quantum machine learning.
	\newblock {\em Nature}, 549(7671):195--202, 2017.
	
	\bibitem{farhi2014qaoa}
	Edward Farhi, Jeffrey Goldstone, and Sam Gutmann.
	\newblock A quantum approximate optimization algorithm, 2014.
	
	\bibitem{larocca2022groupinvqml}
	Mart\'{\i}n Larocca, Fr\'ed\'eric Sauvage, Faris~M. Sbahi, Guillaume Verdon,
	Patrick~J. Coles, and M.~Cerezo.
	\newblock Group-invariant quantum machine learning.
	\newblock {\em PRX Quantum}, 3:030341, Sep 2022.
	
	\bibitem{lins24}
	Isis~Didier Lins, Lavínia Maria~Mendes Araújo, Caio Bezerra~Souto Maior,
	Plínio~Marcio da~Silva~Ramos, Márcio~José das Chagas~Moura, André~Juan
	Ferreira-Martins, Rafael Chaves, and Askery Canabarro.
	\newblock Quantum machine learning for drowsiness detection with eeg signals.
	\newblock {\em Process Safety and Environmental Protection}, 186:1197--1213,
	2024.
	
	\bibitem{Mcclean_nature_VQE_photonic}
	Alberto Peruzzo, Jarrod McClean, Peter Shadbolt, Man-Hong Yung, Xiao-Qi Zhou,
	Peter~J. Love, Al{\'{a}}n Aspuru-Guzik, and Jeremy~L. O'Brien.
	\newblock A variational eigenvalue solver on a photonic quantum processor.
	\newblock {\em Nature Communications}, 5(1), July 2014.
	
	\bibitem{tilly2022variational}
	Jules Tilly, Hongxiang Chen, Shuxiang Cao, Dario Picozzi, Kanav Setia, Ying Li,
	Edward Grant, Leonard Wossnig, Ivan Rungger, George~H Booth, et~al.
	\newblock The variational quantum eigensolver: a review of methods and best
	practices.
	\newblock {\em Physics Reports}, 986:1--128, 2022.
	
	\bibitem{cerezo2022stateeigensolver}
	M~Cerezo, Kunal Sharma, Andrew Arrasmith, and Patrick~J Coles.
	\newblock Variational quantum state eigensolver.
	\newblock {\em npj Quantum Information}, 8(1):113, 2022.
	
	\bibitem{higgott2019variational}
	Oscar Higgott, Daochen Wang, and Stephen Brierley.
	\newblock Variational quantum computation of excited states.
	\newblock {\em Quantum}, 3:156, 2019.
	
	\bibitem{jones2019variational}
	Tyson Jones, Suguru Endo, Sam McArdle, Xiao Yuan, and Simon~C Benjamin.
	\newblock Variational quantum algorithms for discovering hamiltonian spectra.
	\newblock {\em Physical Review A}, 99(6):062304, 2019.
	
	\bibitem{liu2019vqe}
	Jin-Guo Liu, Yi-Hong Zhang, Yuan Wan, and Lei Wang.
	\newblock Variational quantum eigensolver with fewer qubits.
	\newblock {\em Phys. Rev. Res.}, 1:023025, Sep 2019.
	
	\bibitem{kardashin2020certified}
	Andrey Kardashin, Alexey Uvarov, Dmitry Yudin, and Jacob Biamonte.
	\newblock Certified variational quantum algorithms for eigenstate preparation.
	\newblock {\em Physical Review A}, 102(5):052610, 2020.
	
	\bibitem{consiglio2022variational}
	Mirko Consiglio, Tony~JG Apollaro, and Marcin Wie{\'s}niak.
	\newblock Variational approach to the quantum separability problem.
	\newblock {\em Physical Review A}, 106(6):062413, 2022.
	
	\bibitem{plenio2005introduction}
	Martin~B Plenio and Shashank Virmani.
	\newblock An introduction to entanglement measures.
	\newblock {\em arXiv preprint quant-ph/0504163}, 2005.
	
	\bibitem{horodecki2009quantum}
	Ryszard Horodecki, Pawe{\l} Horodecki, Micha{\l} Horodecki, and Karol
	Horodecki.
	\newblock Quantum entanglement.
	\newblock {\em Reviews of modern physics}, 81(2):865, 2009.
	
	\bibitem{kok2004quantum}
	Pieter Kok, Samuel~L Braunstein, and Jonathan~P Dowling.
	\newblock Quantum lithography, entanglement and heisenberg-limited parameter
	estimation.
	\newblock {\em Journal of Optics B: Quantum and Semiclassical Optics},
	6(8):S811, 2004.
	
	\bibitem{bruss2002optimal}
	Dagmar Bruss and Chiara Macchiavello.
	\newblock Optimal eavesdropping in cryptography with three-dimensional quantum
	states.
	\newblock {\em Physical review letters}, 88(12):127901, 2002.
	
	\bibitem{terhal2000bell}
	Barbara~M Terhal.
	\newblock Bell inequalities and the separability criterion.
	\newblock {\em Physics Letters A}, 271(5-6):319--326, 2000.
	
	\bibitem{horodecki1996teleportation}
	Ryszard Horodecki, Micha{\l} Horodecki, and Pawe{\l} Horodecki.
	\newblock Teleportation, bell's inequalities and inseparability.
	\newblock {\em Physics Letters A}, 222(1-2):21--25, 1996.
	
	\bibitem{friis2019entanglement}
	Nicolai Friis, Giuseppe Vitagliano, Mehul Malik, and Marcus Huber.
	\newblock Entanglement certification from theory to experiment.
	\newblock {\em Nature Reviews Physics}, 1(1):72--87, 2019.
	
	\bibitem{shahandeh2014structural}
	F~Shahandeh, J~Sperling, and W~Vogel.
	\newblock Structural quantification of entanglement.
	\newblock {\em Physical review letters}, 113(26):260502, 2014.
	
	\bibitem{weilenmann2020entanglement}
	Mirjam Weilenmann, Benjamin Dive, David Trillo, Edgar~A Aguilar, and Miguel
	Navascu{\'e}s.
	\newblock Entanglement detection beyond measuring fidelities.
	\newblock {\em Physical Review Letters}, 124(20):200502, 2020.
	
	\bibitem{amaro2020design}
	David Amaro and Markus M{\"u}ller.
	\newblock Design and experimental performance of local entanglement witness
	operators.
	\newblock {\em Physical Review A}, 101(1):012317, 2020.
	
	\bibitem{eisert2007quantitative}
	Jens Eisert, Fernando~GSL Brandao, and Koenraad~MR Audenaert.
	\newblock Quantitative entanglement witnesses.
	\newblock {\em New Journal of Physics}, 9(3):46, 2007.
	
	\bibitem{wang2024probing}
	Ke~Wang, Weikang Li, Shibo Xu, Mengyao Hu, Jiachen Chen, Yaozu Wu, Chuanyu
	Zhang, Feitong Jin, Xuhao Zhu, Yu~Gao, et~al.
	\newblock Probing many-body bell correlation depth with superconducting qubits.
	\newblock {\em arXiv preprint arXiv:2406.17841}, 2024.
	
	\bibitem{dowling2004energy}
	Mark~R Dowling, Andrew~C Doherty, and Stephen~D Bartlett.
	\newblock Energy as an entanglement witness for quantum many-body systems.
	\newblock {\em Physical Review A}, 70(6):062113, 2004.
	
	\bibitem{guhne2009entanglement}
	Otfried G{\"u}hne and G{\'e}za T{\'o}th.
	\newblock Entanglement detection.
	\newblock {\em Physics Reports}, 474(1-6):1--75, 2009.
	
	\bibitem{yamamoto1998elementary}
	Shoji Yamamoto, S~Brehmer, and H-J Mikeska.
	\newblock Elementary excitations of heisenberg ferrimagnetic spin chains.
	\newblock {\em Physical Review B}, 57(21):13610, 1998.
	
	\bibitem{vznidarivc2008many}
	Marko {\v{Z}}nidari{\v{c}}, Toma{\v{z}} Prosen, and Peter Prelov{\v{s}}ek.
	\newblock Many-body localization in the heisenberg x x z magnet in a random
	field.
	\newblock {\em Physical Review B}, 77(6):064426, 2008.
	
	\bibitem{amico2008entanglement}
	Luigi Amico, Rosario Fazio, Andreas Osterloh, and Vlatko Vedral.
	\newblock Entanglement in many-body systems.
	\newblock {\em Reviews of modern physics}, 80(2):517, 2008.
	
	\bibitem{le2018spin}
	Thao~P Le, Jesper Levinsen, Kavan Modi, Meera~M Parish, and Felix~A Pollock.
	\newblock Spin-chain model of a many-body quantum battery.
	\newblock {\em Physical Review A}, 97(2):022106, 2018.
	
	\bibitem{bosse2022probing}
	Jan~Lukas Bosse and Ashley Montanaro.
	\newblock Probing ground-state properties of the kagome antiferromagnetic
	heisenberg model using the variational quantum eigensolver.
	\newblock {\em Physical Review B}, 105(9):094409, 2022.
	
	\bibitem{kattemolle2022variational}
	Joris Kattem{\"o}lle and Jasper Van~Wezel.
	\newblock Variational quantum eigensolver for the heisenberg antiferromagnet on
	the kagome lattice.
	\newblock {\em Physical Review B}, 106(21):214429, 2022.
	
	\bibitem{Learning_qearthmover_cost}
	Bobak~Toussi Kiani, Giacomo~De Palma, Milad Marvian, Zi-Wen Liu, and Seth
	Lloyd.
	\newblock Learning quantum data with the quantum earth mover’s distance.
	\newblock {\em Quantum Science and Technology}, 7(4):045002, jul 2022.
	
	\bibitem{costfunction_tracedistance_operators}
	Sumeet Khatri, Ryan LaRose, Alexander Poremba, Lukasz Cincio, Andrew~T.
	Sornborger, and Patrick~J. Coles.
	\newblock Quantum-assisted quantum compiling.
	\newblock {\em {Quantum}}, 3:140, May 2019.
	
	\bibitem{benchmarking_eigensolvers}
	Jiaqi Hu, Junning Li, Yanling Lin, Hanlin Long, Xu-Sheng Xu, Zhaofeng Su,
	Wengang Zhang, Yikang Zhu, and Man-Hong Yung.
	\newblock Benchmarking variational quantum eigensolvers for quantum chemistry,
	2022.
	
	\bibitem{dallaire2020application}
	Pierre-Luc Dallaire-Demers, Micha{\l} St{\k{e}}ch{\l}y, Jerome~F Gonthier,
	Ntwali~Toussaint Bashige, Jonathan Romero, and Yudong Cao.
	\newblock An application benchmark for fermionic quantum simulations.
	\newblock {\em arXiv preprint arXiv:2003.01862}, 2020.
	
	\bibitem{symmetryadapted_ansatz}
	Kazuhiro Seki, Tomonori Shirakawa, and Seiji Yunoki.
	\newblock Symmetry-adapted variational quantum eigensolver.
	\newblock {\em Phys. Rev. A}, 101:052340, May 2020.
	
	\bibitem{buildingspaticalsymmetries_ansatz}
	Frederic Sauvage, Martin Larocca, Patrick~J. Coles, and M.~Cerezo.
	\newblock Building spatial symmetries into parameterized quantum circuits for
	faster training, 2022.
	
	\bibitem{benchmarking_quantumoptimalcontrol}
	Alexandre Choquette, Agustin Di~Paolo, Panagiotis~Kl. Barkoutsos, David
	S\'en\'echal, Ivano Tavernelli, and Alexandre Blais.
	\newblock Quantum-optimal-control-inspired ansatz for variational quantum
	algorithms.
	\newblock {\em Phys. Rev. Res.}, 3:023092, May 2021.
	
	\bibitem{choquette2021quantum}
	Alexandre Choquette, Agustin Di~Paolo, Panagiotis~Kl Barkoutsos, David
	S{\'e}n{\'e}chal, Ivano Tavernelli, and Alexandre Blais.
	\newblock Quantum-optimal-control-inspired ansatz for variational quantum
	algorithms.
	\newblock {\em Physical Review Research}, 3(2):023092, 2021.
	
	\bibitem{nielsenchuangbook}
	Michael~A. Nielsen and Isaac~L. Chuang.
	\newblock {\em Quantum Computation and Quantum Information: 10th Anniversary
		Edition}.
	\newblock Cambridge University Press, 2011.
	
	\bibitem{arute2019quantum}
	Frank Arute, Kunal Arya, Ryan Babbush, Dave Bacon, Joseph~C Bardin, Rami
	Barends, Rupak Biswas, Sergio Boixo, Fernando~GSL Brandao, David~A Buell,
	et~al.
	\newblock Quantum supremacy using a programmable superconducting processor.
	\newblock {\em Nature}, 574(7779):505--510, 2019.
	
	\bibitem{katabarwa2022connecting}
	Amara Katabarwa, Sukin Sim, Dax~Enshan Koh, and Pierre-Luc Dallaire-Demers.
	\newblock Connecting geometry and performance of two-qubit parameterized
	quantum circuits.
	\newblock {\em Quantum}, 6:782, 2022.
	
	\bibitem{wu2021efficient}
	Anbang Wu, Gushu Li, Yuke Wang, Boyuan Feng, Yufei Ding, and Yuan Xie.
	\newblock Towards efficient ansatz architecture for variational quantum
	algorithms, 2021.
	
	\bibitem{zhou2020quantum}
	Leo Zhou, Sheng-Tao Wang, Soonwon Choi, Hannes Pichler, and Mikhail~D Lukin.
	\newblock Quantum approximate optimization algorithm: Performance, mechanism,
	and implementation on near-term devices.
	\newblock {\em Physical Review X}, 10(2):021067, 2020.
	
	\bibitem{HVA_refadicional}
	Wen~Wei Ho and Timothy~H. Hsieh.
	\newblock Efficient variational simulation of non-trivial quantum states.
	\newblock {\em SciPost Phys.}, 6:29, 2019.
	
	\bibitem{dallaire2019low}
	Pierre-Luc Dallaire-Demers, Jonathan Romero, Libor Veis, Sukin Sim, and
	Al{\'a}n Aspuru-Guzik.
	\newblock Low-depth circuit ansatz for preparing correlated fermionic states on
	a quantum computer.
	\newblock {\em Quantum Science and Technology}, 4(4):045005, 2019.
	
	\bibitem{bergholm2018pennylane}
	Ville Bergholm, Josh Izaac, Maria Schuld, Christian Gogolin, Shahnawaz Ahmed,
	Vishnu Ajith, M~Sohaib Alam, Guillermo Alonso-Linaje, B~AkashNarayanan, Ali
	Asadi, et~al.
	\newblock Pennylane: Automatic differentiation of hybrid quantum-classical
	computations.
	\newblock {\em arXiv preprint arXiv:1811.04968}, 2018.
	
	\bibitem{correr2024characterizing}
	Guilherme~Il{\'a}rio Correr, Ivan Medina, Pedro~C Azado, Alexandre Drinko, and
	Diogo~O Soares-Pinto.
	\newblock Characterizing randomness in parameterized quantum circuits through
	expressibility and average entanglement.
	\newblock {\em arXiv preprint arXiv:2405.02265}, 2024.
	
\end{thebibliography}

\end{document}